\documentclass[showpacs,amsmath,amssymb,aps,showkeys,floatfix,prd,a4paper]{revtex4}
\usepackage[utf8]{inputenc} 
\usepackage{verbatim} 
\usepackage[dvips]{graphicx}
\usepackage{dcolumn}
\usepackage{bm}
\usepackage{epsfig}
\usepackage{amsfonts}
\usepackage{amssymb,amscd}
\def\lsim{\raise0.3ex\hbox{$<$\kern-0.75em\raise-1.1ex\hbox{$\sim$}}}
\def\gsim{\raise0.3ex\hbox{$>$\kern-0.75em\raise-1.1ex\hbox{$\sim$}}}

\newcommand{\rd}{\mbox{\boldmath $\Delta$}}
\newcommand{\ba}{\begin{eqnarray}}

\newcommand{\rr}{\mbox{\boldmath $r$}}

\newcommand{\rb}{\mbox{\boldmath $b$}}
\def\gappeq{\mathrel{\rlap {\raise.5ex\hbox{$>$}}
{\lower.5ex\hbox{$\sim$}}}}
\def\lappeq{\mathrel{\rlap{\raise.5ex\hbox{$<$}}
{\lower.5ex\hbox{$\sim$}}}}
\def\Toprel#1\over#2{\mathrel{\mathop{#2}\limits^{#1}}}

\begin{document}

\title{Exclusive vector meson photoproduction with a leading baryon in  
photon - hadron interactions at hadronic colliders}
\author{F. Carvalho$^{1}$, V.P. Gon\c{c}alves$^{2}$, F.S. Navarra$^{3,4}$ and D. Spiering$^{3}$}
\affiliation{
$^{1}$ Universidade Federal de S\~ao Paulo\\
CEP 01302-907, S\~{a}o Paulo, Brazil.\\
$^{2}$ High and Medium Energy Group, Instituto de F\'{\i}sica 
e Matem\'atica,  Universidade Federal de Pelotas\\
Caixa Postal 354,  96010-900, Pelotas, RS, Brazil.\\
$^3$Instituto de F\'{\i}sica, Universidade de S\~{a}o Paulo,
C.P. 66318,  05315-970 S\~{a}o Paulo, SP, Brazil.\\
$^4$ Institut de Physique Th\'eorique, Universit\'e Paris Saclay,\\
CEA, CNRS, F-91191, Gif-sur-Yvette, France.
}

\begin{abstract}
Exclusive vector meson photoproduction associated with a                
leading baryon ($B = n, \Delta^+, \Delta^0$) in $pp$ and $pA$ collisions at 
RHIC and LHC energies is investigated using the color dipole formalism and  
taking into account nonlinear effects in the QCD dynamics. In particular, we 
compute the cross sections for $\rho$, $\phi$ and $J/\Psi$ production together 
with a $\Delta$ and compare the predictions with those obtained for a   
leading neutron. Our results show that the $V + \Delta$ cross section is almost   
 30 \% of the $V + n$ one. Our results also show that a future experimental 
analysis of these processes is, in principle, feasible and can be useful to 
study  leading particle production. 

\end{abstract}

\pacs{12.38.-t, 24.85.+p, 25.30.-c}

\keywords{Quantum Chromodynamics, Exclusive vector meson production, Leading 
baryon processes, Saturation effects.}

\maketitle

A good description of  particle production at forward rapidities and high energies is 
fundamental to our understanding of Collider and Cosmic Ray Physics \cite{FP}.      
Recent results indicate that an accurate knowledge of the leading particle momentum 
spectrum  and its energy dependence is crucial for the interpretation of  cosmic 
ray data (See e.g. Ref. \cite{cr}). Additionally, as at forward rapidities and high 
energies we probe the small Bjorken - $x$ components of the target wave  
function, particle   
production in this kinematical range is directly connected to     
the QCD non-linear  dynamics at high energies \cite{hdqcd}. 

The sucessful operation of the HERA $ep$ collider at DESY from 1991 to 2007 and, 
more recently, of the hadronic colliders RHIC and LHC  greatly helped us to 
improve our understanding of many aspects of QCD dynamics. However, as several 
questions  remain without answer, these experimental studies must be continued and 
more observables must be investigated. One of the most promising observables to 
constrain  QCD dynamics at high energies is  exclusive vector meson 
photoproduction (EVMP) in hadronic collisions 
\cite{gluon,outros_klein,vicmag_mesons1, 
outros_vicmag_mesons,outros_frankfurt,Schafer,vicmag_update,gluon2,motyka_watt, 
Lappi,griep,Guzey,Martin,glauber1,bruno1,Xie,bruno3,vicnavdiego,tuchin,bruno4}.  
This process is characterized by two rapidity gaps and two intact hadrons in the    
final state, with the cross sections being proportional to the square of the target 
gluon distribution (in leading logarithmic approximation \cite{gluon}) or, 
equivalently, to the square of the dipole - target forward scattering amplitude in 
the color dipole formalism  \cite{vicmag_mesons1}. 
Consequently, EVMP is strongly sensitive to non-linear effects associated to the 
high gluonic density in the target, which are expected to contribute significantly 
to the QCD dynamics at high energies \cite{hdqcd}.

In collisions involving a proton, processes in which the proton dissociates  
(becoming a neutron, for example) are very important. In $ep$ collisions they 
can significantly affect EVMP. In addition to the main reaction 
$e + p \to e  +  V + p$, a non-negligible fraction of vector mesons $V$ 
may come from the reaction with  proton dissociation $e p \to e + V + X$. In the latter reaction 
the proton dissociation would reduce the rapidity gap expected in the former reaction. We are
thus facing two important  challenges: the experimental identification of 
EVMP \cite{FP} (especially for the Run 2 LHC energies due  to the large pileup) 
and the quantitative estimate of the contribution of EVMP  with 
 the proton dissociation \cite{wolf,vicwer}. 
Both subjects are intrinsically related. In order to identify exclusive processes 
without the measurement of  rapidity gaps it is necessary to tag the protons in 
the final state. If the proton dissociates, this   
signature of the event will be destroyed. Therefore, more detailed studies of proton  
dissociative process and/or alternative final states that can be used to tag the 
EVPM are important and timely.          

We have recently studied \cite{LN_upc} one of the possible proton  
dissociation processes, where the proton dissociates into a leading neutron and a 
pion, with the former carrying a large fraction of the proton momentum.
In principle, the presence of a leading neutron in  EVMP
can be used to tag the event using the Zero Degree    
Calorimeters (ZDC) already installed in several of the colliders detectors   
\cite{jesus}. Our results indicated that the associated cross sections are         
non-negligible and that an experimental analysis is feasible. 
One of the advantages  of the approach proposed in Ref. \cite{LN_upc} is that 
it has a strong predictive power. 
It is based on the same assumptions used to sucessfully describe   
the EVMP without proton dissociation \cite{bruno3} 
as well as  inclusive and exclusive $\gamma p$ interactions with a leading neutron 
at HERA \cite{nosLN,nosLN2}. Therefore, the experimental measurement of this 
process   
will impose stringent constraints on the theoretical description of leading 
neutron and of  EVMP as well. 

In this work we will extend and complement our previous study  
\cite{LN_upc} on  EVMP and consider processes where the proton splits into 
$\Delta \pi$ states. These are, after $p \to n \pi$, the next most important 
proton dissociation process and their influence on the leading neutron longitudinal 
momentum ($x_L$) spectrum  measured at HERA has been investigated by the H1 and ZEUS 
collaborations with the 
help of their standard event generators. The conclusion, presented in Ref. 
\cite{schmidke}, is that the presence of $\Delta$ intermediate states leads 
to significantly softer leading neutron spectra. In fact, neutrons coming from 
the process $p \to n \pi$ peak at $x_L \simeq 0.7$ and neutrons coming from the 
$p \to \Delta \pi \to n \pi \pi$ sequential dissociation peak at $ x_L < 0.5$. 
In an analogous calculation, here we compute the rapidity      
distributions and total cross sections for  $\rho$, $\phi$ and $J/\Psi$ production  
associated with a leading baryon ($B =  \Delta^+, \Delta^0$) in $pp$ and $pA$   
collisions at RHIC and LHC energies.

\begin{figure}[t]
    {\psfig{figure=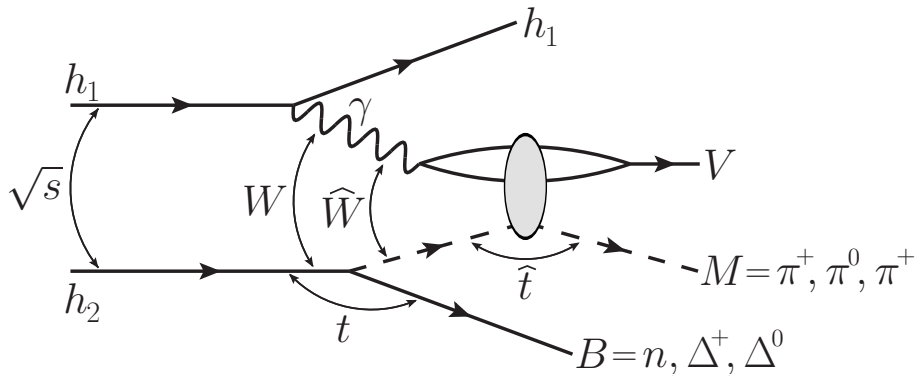,scale=0.65}} 
  \caption{Typical diagram of exclusive vector meson photoproduction in 
association with  a leading baryon.}
  \label{Fig:diagrama}
\end{figure}

We start our analysis presenting a brief review of the formalism needed to 
describe  EVMP  associated with a leading particle in     
photon-induced interactions at hadronic collisions. The process is represented 
in Fig. \ref{Fig:diagrama} for a generic hadronic $h_1 h_2$ collision considering 
that a leading baryon is produced in association with the vector meson.  It will be 
characterized by one rapidity gap associated to the photon exchange and another 
between the vector meson and the meson $M$ due to the diffractive interaction. 
The basic assumption in the description of photon - induced interactions at  
hadronic colliders is that the corresponding hadronic cross sections  can be  
factorized in terms of the equivalent flux of photons and the photon-target cross   
section. As a consequence, the rapidity distribution of mesons produced in  EVMP 
in association with a leading baryon  is given by \cite{LN_upc}
\begin{eqnarray}
  \frac{d\sigma \,\left[h_1 + h_2 \rightarrow   h_3 \otimes V  \otimes \pi +      
B \right]}{dY} = \left[\omega 
  \frac{dN}{d\omega}\bigg|_{h_1}\,\sigma_{\gamma h_2 \rightarrow V  \otimes \pi +  
B}\left(\omega \right)\right]_{\omega_L} + 
  \left[\omega \frac{dN}{d\omega}\bigg|_{h_2}\,\sigma_{\gamma h_1 \rightarrow V  
\otimes \pi + B}
  \left(\omega \right)\right]_{\omega_R}\,
  \label{dsigdy}
\end{eqnarray}
where $h_3$ corresponds to the initial hadron ($h_1$ or $h_2$) which has emitted 
the photon,  $Y=\ln(2\omega/M_V)$ is the rapidity of a vector meson with mass 
$M_V$ and $(dN/d\omega)_h$ denotes the  equivalent photon 
spectrum  of the  incident hadron $h$, with the flux of a nucleus    
being enhanced by a factor $Z^2$ in comparison to the proton one. As in Refs.    
\cite{bruno3,LN_upc} we will assume that the photon flux associated to the proton 
and nucleus can be described by  the Drees - Zeppenfeld  \cite{Dress}      
and the relativistic point -- like charge \cite{upc} models, respectively.       
Moreover,  the symbol $\otimes$ in Eq. (\ref{dsigdy}) represents  a rapidity gap    
in the final state and $\omega_L \, (\propto e^{-Y})$ 
and $\omega_R \, (\propto e^{Y})$ are energies of the photons emitted by  
the $h_1$ and $h_2$ hadrons, respectively.  Finally, the photon-target cross 
section is given by  $\sigma_{\gamma h}$,
which depends on the photon energy $\omega$ in the collider frame. Following   
Refs. \cite{nosLN2,LN_upc}, we will describe  the photon-target interaction in    
terms of the  pion splitting function and of the color dipole scattering 
amplitude, such that the photon-target interaction can be seen as a sequence of 
three factorizable subprocesses:
i) the photon emitted by one of the hadrons fluctuates into a quark-antiquark      
pair (the color dipole), 
ii) the color dipole interacts diffractively with the pion emitted by the proton   
(the other hadron), and 
iii) the vector meson and the leading particle are formed. Moreover, as in          
Refs. \cite{nosLN,nosLN2} we will assume that the absorptive corrections associated   
to soft rescatterings \cite{speth,pirner} 
can be approximated by a constant factor $\cal{K}$. The total cross section for the 
process 
$\gamma p \rightarrow V  \otimes \pi + B$ can be expressed by
\begin{eqnarray}
  \sigma_{\gamma p \rightarrow V  \otimes \pi + B} (W^2) = {\cal{K}} \cdot \int dx_L 
dt \, f_{\pi/p} (x_L,t) \cdot 
  \sigma_{\gamma \pi \rightarrow V  \otimes \pi}({\hat{W}}^2)
\label{crossgen}
\end{eqnarray}
where  $W=(2 \omega \sqrt{s})^{\frac{1}{2}}$ is the center-of-mass energy of the  
photon-proton system, 
$x_L$ is the proton momentum fraction carried by the leading particle and 
$t$ is the square of the four-momentum of the exchanged pion. 
Moreover, $f_{\pi/p}$ is the flux of virtual pions emitted by the proton and
$\sigma_{\gamma \pi \rightarrow V  \otimes \pi}(\hat{W}^2)$ is the cross section 
of the interaction between the  photon and the pion at center-of-mass energy $\hat{W}$,
which is given by  $\hat{W}^2 = (1-x_L) \, W^2$. 
Following  Refs. \cite{nosLN,nosLN2,LN_upc}, 
we will assume that the pion flux is given by \cite{koepf96}
\begin{eqnarray}
  f_{\pi/p} (x_L,t) = \frac{g_{p \pi B}^2}{16 \pi^2}
   \frac{\mathcal{B}(t,m_p,m_B)}{(t-m_{\pi}^2)^2} \left(1-x_L\right)^{1-2t}  
  \exp \left[ 2b (t-m_{\pi}^2) \right]
  \label{genflux}  
\end{eqnarray}
where $g_{p \pi B}$ is the proton-pion-baryon coupling constant,
$m_{\pi}$ is the pion mass 
and $b = 0.3$ GeV$^{-2}$ is related to the $p \pi B$ form factor.
The term $\mathcal{B}$ depends of the produced baryon: 
\begin{eqnarray}
  \mathcal{B}(t,m_p,m_B) = 
    \left\{ 
     \begin{array}{ll}
      -t + \left(m_n-m_p\right)^2\,, & \hspace{0.2cm}\text{for $B\!=\!n$,}\\
      \displaystyle\frac{\left[\left(m_{\Delta}+m_p\right)^2-t\right]^2\left[
\left(m_{\Delta}-m_p\right)^2-t\right]}{12\,m_p^2\,m_{\Delta}^2}\,,
       & \hspace{0.2cm}\text{for $B\!=\!\Delta$,}
     \end{array}
    \right.
\end{eqnarray}
where $m_p$, $m_n$ and $m_{\Delta}$ are the respective masses of the proton, 
neutron and delta. In our analysis we will assume that 
 $g_{p\pi^+n} = 19.025$, $g_{p\pi^+\Delta^0} = 11.676$ and $
  g_{p\pi^0\Delta^+} =  16.512$ \cite{babi}. 
Additionally, the 
 $\gamma \pi \rightarrow V  \otimes \pi$ cross section will be expressed  by 
\begin{eqnarray}
  \sigma (\gamma \pi \rightarrow V \otimes \pi) =  \int_{-\infty}^0 \frac{d\sigma}
{d\hat{t}}\, d\hat{t}  
  = \frac{1}{16\pi}  \int_{-\infty}^0 |{\cal{A}}^{\gamma \pi \rightarrow V \pi }
(\hat x,\Delta)|^2 \, d\hat{t}\,\,,
  \label{sctotal_intt}
\end{eqnarray}
where the scattering amplitude is given in the dipole formalism by
\begin{eqnarray}
   {\cal A}^{\gamma \pi \rightarrow V \pi}(\hat{x},\Delta)  =  i
   \int d\alpha \, d^2\rr \, d^2\rb\, e^{-i[\rb-(1-\alpha)\rr]\cdot \rd} 
   \,\, (\Psi^{V*}\Psi)  \,\,2 {\cal{N}}_\pi(\hat{x},\rr,\rb)
   \label{sigmatot2}
\end{eqnarray}
with   $\hat x=M_V^2/\hat W^2$ being the scaled Bjorken variable,
$\hat{t}=-\Delta^2$ denotes the transverse momentum lost by the outgoing pion,   
the variable  $\alpha$ $(1-\alpha)$ is the longitudinal momentum fraction of the    
quark (antiquark) 
whereas the variable $\rb$ is the transverse distance from the center of the target 
to the center of mass of the $q \bar{q}$  dipole.
Finally, 
$(\Psi^{V*}\Psi)$ denotes the overlap between the real photon and exclusive final 
state wave functions, 
which we assume to be given by the Gauss-LC model described in Ref. \cite{nosLN}, 
and  $\mathcal{N}^\pi(\hat{x},\rr,\rb)$ is the imaginary part of the 
forward amplitude of the scattering between a small dipole                          
(a colorless quark-antiquark pair) and a pion, at a given rapidity interval, which  
is directly related to the QCD dynamics at high energies \cite{hdqcd}. 
As in Ref. \cite{LN_upc}, we will assume that $\mathcal{N}^\pi$ 
can be expressed in terms of the dipole-proton scattering amplitude $\mathcal{N}^p$, 
usually probed in the  inclusive and exclusive processes at HERA, as follows 
\begin{eqnarray}
  {\cal N}^\pi (\hat{x}, \rr, \rb) = R_q \cdot {\cal N}^p (\hat{x}, \rr, \rb) \,\,,
  \label{doister}
\end{eqnarray}
with $R_q$ being a constant. Moreover, we  will assume that   ${\cal{N}}^p    
(\hat{x},\rr,\rb)$ is given by the bCGC model proposed in Ref. \cite{kmw} and      
recently updated in Ref. \cite{amir}. It is important to emphasize that this model  
reproduces quite well    the HERA data on exclusive $\rho$ and $J/\Psi$ production   
\cite{amir_armesto} as well as the EVMP  without 
proton dissociation in hadronic collisions \cite{bruno3}.                     
As the formalism that will be used in our study of the EVMP 
associated with a leading baryon is the same extensively discussed in  
our previous papers \cite{nosLN,nosLN2,LN_upc} for the leading neutron production,   
we refer the reader to these references for a more detailed analysis about the       
dependence of our predictions on the meson flux and on the  photon - hadron 
scattering amplitude.

In order to estimate the rapidity distributions and total cross sections for 
EVMP associated with a leading baryon 
we need to specify a model for the absorptive corrections, represented by the 
${\cal{K}}$ 
factor in Eq. (\ref{crossgen}), as well as a value for the $R_q$ factor in Eq.   
(\ref{doister}). The range of possible values for ${\cal{K}}$ was fixed in     
Ref. \cite{nosLN2} using HERA data \cite{rhoLN_HERA} for the $\sigma (\gamma p   
\rightarrow \rho \otimes \pi + n)$ process, 
being given by $({\cal{K}}_{min}, {\cal{K}}_{med}, {\cal{K}}_{max}) =           
(0.152, 0.179, 0.205)$. In what follows we will perform our calculations of the   
rapidity distributions  and total cross sections assuming that $ {\cal{K}}      
= {\cal{K}}_{med}$.  Moreover, as in Refs. \cite{nosLN2,LN_upc}, we will assume   
that $R_q = 2/3$, 
as expected from the additive quark model, which allows to describe the inclusive  
and exclusive HERA data for leading neutron production. As a consequence of these 
assumptions, there is no parameter to be fixed and  we can make  predictions 
which can be confronted with data.

\begin{figure}[t]
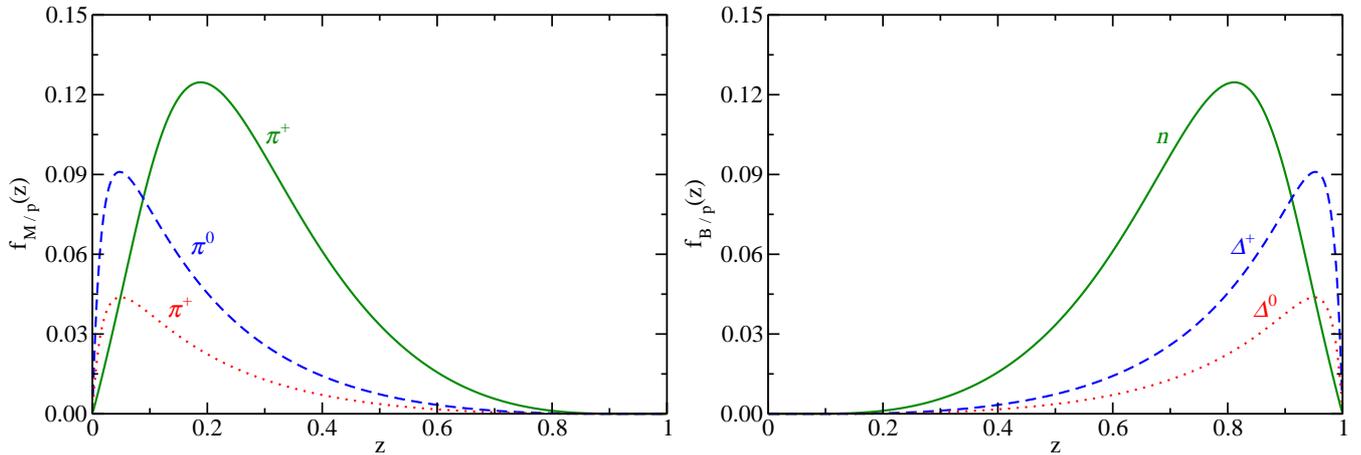

  \begin{tabular}{cc}
    {\psfig{figure=f_meson.eps,scale=0.35}} 
    & {\psfig{figure=f_barion.eps,scale=0.35}}
  \end{tabular}
  \caption{Distribution of the proton fractional momentum ($z$) carried by the 
meson (left)  and by the baryon (right).}
  \label{Fig:fluxes}
\end{figure}


Initially, let us analyze the behavior of the meson  (baryon) flux as a function  
of the fractional momentum  $z$ which the meson (baryon) takes away from the proton.    
The results are presented in Fig. \ref{Fig:fluxes}. As it can be seen, baryons carry  
the largest fraction of the proton momentum, whereas pions populate the  low $z$   
region. Due to energy-momentum conservation,  the position of the peak of the meson 
distribution depends on the  distribution of the associated baryon.
In particular, the pion associated with the neutron has its maximum at  
$z \sim 0.2$,  
while pions associated with the delta states reach their peak at $z \sim 0.05$.   
As a consequence, we expect that a leading 
$\Delta$ carries a larger momentum than a leading neutron. Additionally, 
the typical center - of - mass energies $\hat{W}$ of the photon - meson  
interactions in the case of a leading $\Delta$ will be smaller than those 
in the process with a leading neutron. As it will be seen, this aspect has a 
direct implication on our  predictions for EVMP associated with a leading 
baryon.

\begin{figure}[t]
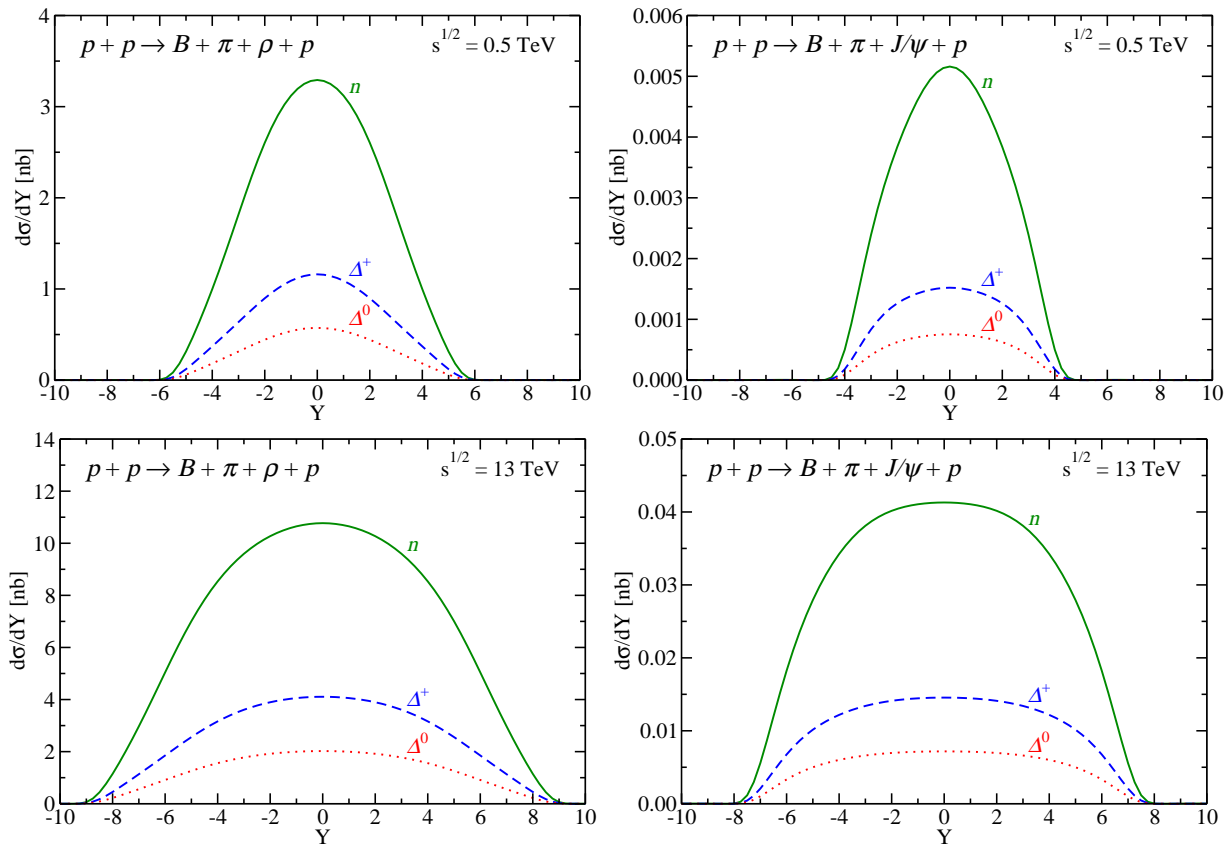

  \begin{tabular}{cc}
   {\psfig{figure=LP_rho_pp_leading_barion_RHIC.eps,scale=0.32}} &
     {\psfig{figure=LP_jpsi_pp_leading_barion_RHIC.eps,scale=0.32}}  \\
    {\psfig{figure=LP_rho_pp_leading_barion_LHC.eps,scale=0.32}} 
    & {\psfig{figure=LP_jpsi_pp_leading_barion_LHC.eps,scale=0.32}} 
  \end{tabular}
  \caption{Vector meson rapidity distributions in EVMP associated with a leading 
baryon in $pp$ collisions. Upper and panels: $\sqrt{s}=$ 0.5 TeV and 13 TeV 
respectively. Left and right panels: $\rho$ and $J/\psi$ respectively.} 
  \label{Fig:pp}
\end{figure}

In Fig. \ref{Fig:pp} we present our predictions for the exclusive  $\rho$ and 
$J/\Psi$ photoproduction associated with a leading baryon in $pp$ collisions at 
RHIC ($\sqrt{s} = 0.5$ TeV)  and LHC ($\sqrt{s} = 13$ TeV)  energies. The 
predictions for the leading $n$, $\Delta^+$ and $\Delta^0$ are presented 
separately. As both incident protons act as photon sources, we have rapidity 
distributions that are symmetric with respect to $Y=0$. The  
predictions for midrapidities $Y \approx 0$ increase with                       
$\sqrt{s}$ and decrease with $M_V$. Additionally, the growth with the energy is    
faster for $J/\Psi$ than for $\rho$ production. Such behavior is expected from the  
non-linear QCD dynamics \cite{hdqcd}, which predicts a larger contribution of these   
effects for processes dominated by larger dipole sizes, as is the case of the $\rho$ 
production in comparison to the $J/\Psi$ one. We predict that the rapidity          
distributions of particles produced in  association with a leading $\Delta$ are 
smaller than those for a leading  
neutron. This difference is directly related to the distinct magnitude of the   
coupling constants and to the different range of center - of - mass photon - meson  
energies probed in the two processes. As already observed in Fig. \ref{Fig:fluxes},  
in the process with a leading neutron the typical values of $z$, and consequently   
$\hat{W}^2$, are larger than those present with a leading $\Delta$. As the      
$\gamma \pi$ cross section increases with the energy, this implies that the cross   
sections for the processes with a leading neutron will be larger. Moreover, we     
observe that the predictions for the leading $\Delta^0$ are smaller than the      
leading $\Delta^+$ by a factor $\sim 2$ at central rapidities, which can be traced   
back to the fact that   $g_{p\,\pi^0\Delta^+} > g_{p\,\pi^+\Delta^0}$.

\begin{figure}[t]
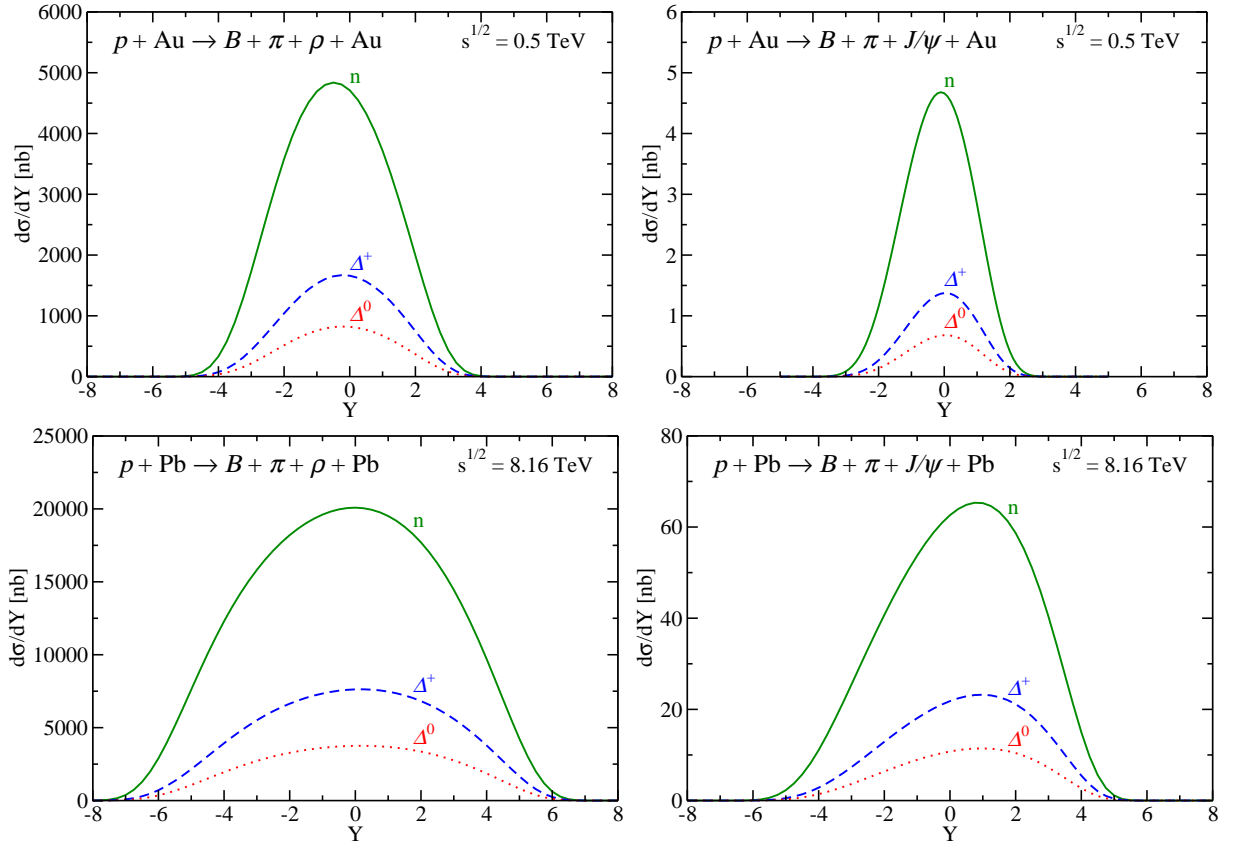

  \begin{tabular}{cc}
   {\psfig{figure=LP_rho_pA_leading_barion_RHIC.eps,scale=0.32}} & 
     {\psfig{figure=LP_jpsi_pA_leading_barion_RHIC.eps,scale=0.32}}  \\
     {\psfig{figure=LP_rho_pA_leading_barion_LHC.eps,scale=0.32}}
     & {\psfig{figure=LP_jpsi_pA_leading_barion_LHC.eps,scale=0.32}}
  \end{tabular}
  \caption{
Vector meson rapidity distributions in EVMP associated with a leading
baryon in $pA$ collisions. Upper and panels: $\sqrt{s}=$ 0.5 TeV and 8.16 TeV
respectively. Left and right panels: $\rho$ and $J/\psi$ respectively.}
  \label{Fig:pA}
\end{figure}

The results for  exclusive  $\rho$ and $J/\Psi$ photoproduction associated with 
a leading baryon in $pA$ collisions at RHIC ($\sqrt{s} = 0.5$ TeV)  and LHC  
($\sqrt{s} = 8.1$ TeV)  energies are presented in Fig. \ref{Fig:pA}. In our  
calculations we have assumed that $A = 197$ (208) for RHIC (LHC).              
As the nuclear photon flux is enhanced by a factor $Z^2$ in comparison to the   
proton one, we obtain asymmetric rapidity distributions. Similar enhancement is  
predicted in the magnitude of the rapidity distributions at $Y \sim 0$. As in     
the $pp$ case, the EVMP associated with  a leading  
neutron is a factor $\gtrsim 3 \, (5)$ larger than those associated to a leading 
$\Delta^+ \,(\Delta^0)$.

\begin{table}[t]
  \centering
  \begin{tabular}{|c|c|c|c|c|c|c|c|c|}
    \hline
    & \multicolumn{3}{c}{\phantom{xxxxxxx}$pp$} &  &\multicolumn{3}{c}
{\phantom{xxxxxxx}$pA$} &\\
    \hline
    VM  &  $\sqrt{s}/$TeV  & $\sigma(n\pi^+)$/nb & 
    $\sigma(\Delta^0\pi^+)$/nb & $\sigma(\Delta^+\pi^0)$/nb & 
    $\sqrt{s}/$TeV  & $\sigma(n\pi^+)$/$\mu$b & 
    $\sigma(\Delta^0\pi^+)$/$\mu$b & $\sigma(\Delta^+\pi^0)$/$\mu$b\\
    \hline
    $\rho$   & 0.2  & 10.41  & 1.74   & 3.53  & 0.2  & 7.71   & 1.18  & 2.39  \\ 
    $\rho$   & 0.5  & 21.00  & 3.64   & 7.39  & 0.5  & 21.20  & 3.47  & 7.03  \\
    $\rho$   & 8.0  & 97.85  & 17.99  & 36.53 & 5.02 & 124.03 & 22.09 & 44.82 \\
    $\rho$   & 13.0 & 121.34 & 22.45  & 45.59 & 8.16 & 163.12 & 29.35 & 59.56 \\
    \hline
    $\phi$   & 0.2  & 1.71   & 0.28   & 0.57  & 0.2  & 1.09   & 0.16   & 0.33   \\
    $\phi$   & 0.5  & 3.65   & 0.63   & 1.27  & 0.5  & 3.37   & 0.54   & 1.09  \\
    $\phi$   & 8.0  & 18.66  & 3.41   & 6.93  & 5.02 & 22.43  & 3.96  & 8.03  \\
    $\phi$   & 13.0 & 23.39  & 4.30   & 8.74  & 8.16 & 29.99  & 5.35  & 10.86 \\
    \hline
    $J/\psi$ & 0.2  & 0.01   & 0.001   & 0.002  & 0.2  & 1.63 $\times 10^{-3}$       
& 0.21 $\times 10^{-3}$     & 0.42 $\times 10^{-3}$     \\                         
    $J/\psi$ & 0.5  & 0.03   & 0.004   & 0.008  & 0.5  & 12.66 $\times 10^{-3}$     
& 1.80 $\times 10^{-3}$     & 3.63 $\times 10^{-3}$     \\
    $J/\psi$ & 8.0  & 0.32   & 0.056   & 0.114  & 5.02 & 0.25    & 0.04    & 0.08    \\
    $J/\psi$ & 13.0 & 0.45   & 0.079   & 0.160  & 8.16 & 0.37    & 0.06    & 0.13   \\
    \hline  
  \end{tabular}
  \caption{Total cross sections of EVMP associated with a leading baryon in 
$pp$ and $pA$ collisions at different center-of-mass energies. In the case of $pA$ 
collisions we consider $A = 197$ (208) for RHIC (LHC) energies.}
  \label{Tab:total_LP}
\end{table}

Let us now estimate the total cross sections for  $pp$ collisions at     
$\sqrt{s} = 0.2, \, 0.5, \, 8 \text{ and } 13$ TeV,  $pAu$ collisions at   
$\sqrt{s} = 0.2 \text{ and } 0.5$ TeV and $pPb$ collisions at 
$\sqrt{s} = 5.02 \text{ and } 8.16$ TeV.                         
As in Refs. \cite{rhoLN_HERA,nosLN2,LN_upc}, we will assume that                
$\sqrt{|t|} <  0.2$ GeV. Moreover, we will also present our predictions for 
exclusive $\phi$ photoproduction.
We have verified that the rapidity distributions for this final state are similar 
to those for the $\rho$ production, but smaller in magnitude.
In Table \ref{Tab:total_LP} we present our predictions. 
As expected from the analysis of the rapidity distributions,                  
the cross sections increase with the energy and decrease with the mass of the      
vector meson. Moreover, their magnitude is enhanced in $pA$ collisions in 
comparison with  $pp$ collisions.
For the exclusive $\rho$ photoproduction associated with a leading neutron, 
we predict values of the order of $10^2$ ($10^5$) nb in $pp$ ($pPb$) collisions     
at LHC energies. 

In the experimental analysis, it may be possible to reconstruct the $\Delta$. 
Then the predictions shown in Figs.~\ref{Fig:pp} and \ref{Fig:pA} can be 
directly compared 
to data. If the reconstruction of $\Delta$ is not possible, then we have to 
remember that the  $\Delta\rightarrow N\pi$ decay channel    
corresponds to 99.4\% of the delta decays \cite{pdg16}. Then the experimentally 
measured  EVPM cross sections and rapidity distributions and also the associated 
leading neutron momentum distributions will receive a significant contribution from 
processes with delta resonances.  Indeed, 
our results indicate that the channels $\Delta^0\rightarrow n\,\pi^0$ and 
$\Delta^+\rightarrow n\,\pi^+$  together have a cross section which is  about  
$20\sim 28$\% of the direct leading neutron production cross section.  The    
secondary $\Delta$ decay into $n \pi$ will not affect the rapidity 
distribution
of the vector meson, but it will give a contribution to the neutron spectrum 
which is  softer than the one coming from the primary $p \to n \, \pi$ splitting.

In comparison with the usual EVMP \cite{bruno3}, EVMP with a leading baryon is 
smaller by approximately two (three) orders of magnitude in the case of a leading 
neutron (delta). However, it is important to emphasize that these 
events will be characterized by  very forward baryons, which can be used to tag 
the events.

To conclude: recent results on 
photon - induced  interactions at hadronic colliders  have demonstrated 
that the analysis of these processes is feasible at  RHIC and LHC, and that it 
is possible to use the resulting experimental data to investigate e.g.  the 
nuclear effects in the gluon distribution, the QCD dynamics at high energies and 
several other issues that still lack  a satisfactory understanding.    

This possibility  has stimulated the improvement of the theoretical description  
of these processes as well as the proposal of complementary processes that also 
probe the QCD dynamics and are more easily tagged in collisions with a high 
pileup. Along this line, we have recently proposed the study of  
EVMP associated with a leading neutron in $\gamma p$ interactions 
at $pp$ and $pA$ collisions and obtained  large values for the total cross sections  
and event rates. This result motivated the analysis performed in the present work, 
where  we have extended the study to other leading particles, with higher mass, 
which also generate a neutron in the final state through its decay. We have  
estimated the exclusive $\rho$, $\phi$ and $J/\Psi$ photoproduction associated 
with a leading neutron and a leading $\Delta$ 
in $pp$ and $pA$ collisions at RHIC and LHC energies. We have found that the  
production associated with a leading $\Delta$  is non-negligible, being about 
30 \% of the one  with a leading neutron. Our results indicate that  the 
experimental analysis of these process is, in principle, feasible. In particular, 
if a combined analysis of the events using central and forward detectors is  
performed, as those expected to occur using the CMS-TOTEM Precision Proton 
Spectrometer \cite{ctpps} and ATLAS + LHCf experiments \cite{lhcf}. 
We expect thus that our results motivate a future experimental analysis of 
EVMP associated with a leading baryon in  
hadronic collisions at RHIC and LHC colliders.

\begin{acknowledgments}
This work was  partially financed by the Brazilian funding agencies CNPq, CAPES, 
FAPERGS, FAPESP (contract number 12/50984-4) and INCT-FNA 
(process number 464898/2014-5). 

\end{acknowledgments}

\hspace{1.0cm}

\end{document}